\documentclass[letterpaper,aps,prl,reprint,twocolumn,superscriptaddress]{revtex4-1}
\usepackage{bm}
\usepackage{color}
\usepackage{graphicx}
\usepackage{amsmath}
\usepackage{amssymb}
\usepackage{amsfonts}
\usepackage{txfonts}
\DeclareMathSizes{10.0}{9.0}{6.5}{6.5} 
\DeclareSymbolFont{operators}{OT1}{cmr}{m}{n}
\DeclareSymbolFont{letters}{OML}{cmm}{m}{it}
\DeclareSymbolFont{symbols}{OMS}{cmsy}{m}{n}
\DeclareSymbolFont{largesymbols}{OMX}{cmex}{m}{n}

\usepackage{subfigure}
\usepackage{epstopdf}
\usepackage[colorlinks=true,citecolor=blue,linkcolor=blue,urlcolor=blue]{hyperref}

\newcommand{\erf}{\mathrm{erf}}
\newcommand\trbf{\sigma}

\begin{document}

\title{The Gaussian Radial Basis Function Method for Plasma Kinetic Theory}
\author{E. Hirvijoki}
\affiliation{Department of Applied Physics, Chalmers University of Technology, 
SE-41296 Gothenburg, SWEDEN}
\author{J. Candy}
\affiliation{General Atomics, PO Box 85608, San Diego, CA 92186-5608, USA}
\author{E. Belli}
\affiliation{General Atomics, PO Box 85608, San Diego, CA 92186-5608, USA}
\author{O. Embr\'eus}
\affiliation{Department of Applied Physics, Chalmers University of Technology, 
SE-41296 Gothenburg, SWEDEN}

\date{\today}

\begin{abstract}  
  A fundamental macroscopic description of a magnetized plasma 
  is the Vlasov equation supplemented by the nonlinear inverse-square force
  Fokker-Planck collision operator [Rosenbluth et al., Phys. Rev., {\bf 107}, 1957].
  The Vlasov part describes advection in a six-dimensional phase space whereas
  the collision operator involves friction and diffusion coefficients that are
  weighted velocity-space integrals of the particle distribution function.
  The Fokker-Planck collision operator is an integro-differential, bilinear
  operator, and numerical discretization of the operator is far from trivial.
  In this letter, we describe a new approach to discretize the entire kinetic
  system based on an expansion in Gaussian Radial Basis functions (RBFs).  This
  approach is particularly well-suited to treat the collision operator because
  the friction and diffusion coefficients can be analytically calculated.  Although
  the RBF method is known to be a powerful scheme for the interpolation of
  scattered multidimensional data, Gaussian RBFs also have a deep physical interpretation as
  local thermodynamic equilibria.  In this letter we outline the general theory,
  highlight the connection to plasma fluid theories, and also give 2D and 3D
  numerical solutions of the nonlinear Fokker-Planck equation.  A broad spectrum
  of applications for the new method is anticipated in both astrophysical and
  laboratory plasmas.
\end{abstract} 


\maketitle 


{\bf Motivation}~~A fundamental macroscopic description of a magnetized plasma is the
Vlasov equation supplemented by the nonlinear inverse-square force Fokker-planck
collision operator \cite{rosenbluth:PhysRev.107.1}
\begin{equation} 
\frac{\partial f_{a}}{\partial t} + \mathbf{v} \cdot \nabla f_{a} +
\frac{e_a}{m_a} \left( \mathbf{E} + \frac{\mathbf{v} \times \mathbf{B}}{c} \right)
\frac{\partial f_{a}}{\partial \mathbf{v}} = \sum_b C_{ab} (f_a,f_b)\;,
\label{eq:maineq}
\end{equation}
where $f_a$ is the distribution of species $a$ with charge $e_a$ and mass $m_a$.
The electric and magnetic fields depend on the dsitribution $f_a$ through Maxwell's
equations.  This model 
assumes a statistical description of Coulomb interaction in the limit of small-angle
scattering, with the changes in $f_a$ due to collisions with species $b$ described by
\begin{equation*}
  C_{ab} = \frac{\partial}{\partial\mathbf{v}}\cdot\left[\mathbf{A}_{ab}f_a
    +\frac{\partial}{\partial\mathbf{v}}\cdot\left(\mathbb{D}_{ab}f_a\right)\right] \;.
\end{equation*}
The friction and diffusion coefficients are given by the expressions 
\begin{equation*}
  \mathbf{A}_{ab}=\;L_{ab}\left(1+\frac{m_a}{m_b}\right)
  \frac{\partial\varphi_b}{\partial\mathbf{v}}\;,
  \qquad
  \mathbb{D}_{ab}=-L^{ab}\frac{\partial^2\psi_b}{\partial\mathbf{v}\partial\mathbf{v}}\;,
\end{equation*}
where $L_{ab}=(e_ae_b/m_a\varepsilon_0)^2\ln\Lambda_{ab}$ and $\ln\Lambda_{ab}$ is the Coulomb logarithm
which respresents a physical cut-off for small-angle collisions.  The Rosenbluth potentials 
appearing in the friction and diffusion coefficients are weighted integrals of the distribution
 function
\begin{align*}
  \varphi_b(\mathbf{x},\mathbf{v},t) =
  &-\frac{1}{4\pi}\int d\mathbf{v}' f_b(\mathbf{x},\mathbf{v}',t)\frac{1}{|\mathbf{v}-\mathbf{v}'|} \; \\
  \psi_b(\mathbf{x},\mathbf{v},t) =
  &-\frac{1}{8\pi}\int d\mathbf{v}' f_b(\mathbf{x},\mathbf{v}',t)|\mathbf{v}-\mathbf{v}'| \; 
\end{align*}
and they satisfy the velocity-space Poisson equations $\nabla^2_{\mathbf{v}}\psi_b=\varphi_b$
and $\nabla^2_{\mathbf{v}}\varphi_b=f_b$.  Expressed in terms of the potential functions,
the Fokker-Planck collision operator is
\begin{equation*}
C_{ab} = L_{ab}\left[\frac{m_a}{m_b}f_af_b+\mu_{ab}
  \frac{\partial\varphi_b}{\partial\mathbf{v}}
  \hskip -1pt \cdot \hskip -1pt
  \frac{\partial f_a}{\partial\mathbf{v}}
  -\frac{\partial^2\psi_b}{\partial\mathbf{v}\partial\mathbf{v}}
  \hskip -1pt : \hskip -1pt
  \frac{\partial^2 f_a}{\partial\mathbf{v}\partial\mathbf{v}}\right] \; ,
\end{equation*}
where $\mu_{ab} = m_a/m_b-1$. A common approach for numerical evaluation of $C_{ab}$ follows a
two-phase method where one first inverts
the velocity-space Laplacian operators and then directly evaluates the collision operator.  
Boundary conditions for the Poisson equations can be obtained by limiting the solution to 
a sub-space and evaluating the expressions at the boundary using a multipole expansion of 
the potentials~\cite{cogent:CTPP201410023},
or by imposing the free-space solution outside the sub-space~\cite{Pataki20117840}.  Another
sophisticated approach is based on fast spectral decomposition as described in
Ref.~\cite{pareschi:2000216:jpc}.  
In this letter, we propose a new approach using a mesh-free shifted-Maxwellian representation 
which is intuitively appealing and straightforward to implement. The solution thus obtained is 
$C^{\infty}$ smooth, extends to $v \rightarrow \infty$, and allows compact representation of 
any interesting macroscopic quantity (number, momentum, energy density, and so on). 


{\bf The Gaussian RBF Method}~~To solve the kinetic equation, Eq.~(\ref{eq:maineq}), we
write the total distribution as a finite sum of shifted Maxwellians
\begin{equation*}
f_{a}(\mathbf{x},\mathbf{v},t)= \sum_i w_{a}^{i}(\mathbf{x},t) \, F_a^i (\mathbf{x},\mathbf{v}) \; ,
\end{equation*}
where each Maxwellian,
$F_a^i = \left(\gamma_a^i/\pi\right)^{3/2}\exp{[-\gamma_a^i(\mathbf{v}-\mathbf{v}_a^i)^2]}$,
is normalized to unity and the weights $w_{a}^{i}$ allowed to evolve in time.  The width
parameters $\gamma_a^i$ and mean velocities $\mathbf{v}_a^i$ can be arbitrary functions
of position.  The shifted Maxwellians are nothing other than Gaussian
{\it Radial Basis functions} (RBFs) which have found numerous applications in applied
mathematics -- in particular for the
construction of neural networks \cite{park:1991}.  For compactness, in what follows we will
retain the spatial dependence of all quantities but will not write the dependence explicitly.
The potential functions then take the form
\begin{equation*}
  \varphi_a(\mathbf{v},t) = \sum_i w_{a}^{i}(t) \, \varphi_a^i(\mathbf{v}) \; ,
  \quad
  \psi_a(\mathbf{v},t) = \sum_i w_{a}^{i}(t) \, \psi_a^i(\mathbf{v}),
\end{equation*}
where the Gaussian RBF potentials
$\varphi_a^i=-\sqrt{\gamma_a^i} \Phi(\sqrt{\gamma_a^i} |\mathbf{v}-\mathbf{v}_a^i|)/(4\pi)$ and
$\psi_a^i=-\sqrt{\gamma_a^i}\Psi(\sqrt{\gamma_a^i} |\mathbf{v}-\mathbf{v}_a^i|)/(8\pi)$ are defined
in terms of the functions $\Phi(s)=\erf(s)/s$ and $\Psi(s)=[s+1/(2s)]\erf(s)+\exp(-s^2)/\sqrt{\pi}$,
where $\erf(s)$ is the error function.  We thus find a simple bilinear expression for the complete
nonlinear operator
\begin{equation*}
C_{ab} = \sum_{k,\ell}w_{a}^{k}(t)\,w_{b}^{\ell}(t)\,C_{ab}^{k\ell}(\mathbf{v})\;,
\end{equation*}
where the Gaussian RBF collision tensor is
\begin{equation*}
C_{ab}^{k\ell} = L_{ab}\left[\frac{m_a}{m_b}F_a^k F_b^{\ell}
  + \mu_{ab} \frac{\partial\varphi_a^k}{\partial\mathbf{v}}
  \hskip -1pt \cdot \hskip -1pt \frac{\partial F_b^\ell}{\partial\mathbf{v}}-
  \frac{\partial^2\psi_a^k}{\partial\mathbf{v}\partial\mathbf{v}}
  \hskip -1pt : \hskip -1pt \frac{\partial^2 F_b^\ell}{\partial\mathbf{v}\partial\mathbf{v}}\right] \; ,
\end{equation*}
such that $C_{aa}^{kk}(\mathbf{v})=0$.
As we have analytical expressions for
$F_a^i(\mathbf{v})$, $\varphi_a^i(\mathbf{v})$, and $\psi_a^i(\mathbf{v})$,
the tensor $C_{ab}^{k\ell}(\mathbf{v})$ is easy to implement and fast to
evaluate at any point in velocity space.  In problems with azimuthal symmetry,
a 2D RBF scheme can be developed with axisymmetric {\em ring}-like
RBF-basis:
\begin{equation*}
  F_i=(\gamma_i/\pi)^{3/2}I_0(2\gamma_iv_{i,\perp}v_{\perp})\,
  e^{-\gamma_i(v_{\parallel}-v_{i,\parallel})^2-\gamma_i(v_{\perp}^2+v_{i,\perp}^2)} \; ,
\end{equation*}
where $I_0$ is the order-zero modified Bessel function of the first kind and $(v_\perp,v_\parallel)$
are the cylindrical velocity-space coordinates. Although explicit expressions for axisymmetric RBF
potentials $\varphi_i$ and $\psi_i$ are not available in a closed form, they can be evaluated
numerically to machine precision at any requested point.


{\bf Collocation Options}~~
We describe two different methods for obtaining an ordinary differential equation for the time
evolution of weights: the {\it Galerkin} and the {\it center-collocation} projections.  In the
Galerkin method, the kinetic equation -- already expanded in the RBF basis -- is multiplied by
each individual basis function and then integrated over the entire domain.  Conversely, in the
center-collocation method, the kinetic equation is evaluated at the center of each RBF.  Both
methods yield $N$ equations for the $N$ RBF weights, $w_a^i$, and result in a
differential equation for the weights that can be written in a matrix form.  For the moment,
let us illustrate the method by considering the spatially homogeneous case with no
electromagnetic fields.  Then, the matrix equation is
\begin{equation}
\label{eq:weight_evolution}
\sum_j M_{ij} \frac{\partial w_a^j}{\partial t}
  = \sum_{k,\ell} w_{a}^{k}(t)\,w_{b}^{\ell}(t)\,C_{ab}^{ik\ell}
  \quad \forall\quad i\in\,1,2,3,\dots,
\end{equation}
In the Galerkin projection (GP), the matrix $M_{ij}$ is symmetric, typically diagonally dominant,
and given by the expression
\begin{equation*}
(M_{ij})_{\rm GP}
    =\left(\frac{\gamma_i\gamma_j}{\pi(\gamma_i+\gamma_j)}\right)
    \exp\left[-\frac{\gamma_i\gamma_j}{\gamma_i+\gamma_j}\left(\mathbf{v}_i-\mathbf{v}_j\right)^2\right]\;,
\end{equation*}
whereas in the center-collocation (CC) method, the matrix $M_{ij}$ is no longer necessarily
symmetric, but can still be dominated by the diagonal components:
\begin{equation*}
(M_{ij})_{\rm CC} \equiv f_j(\mathbf{v}_i)=(\gamma_j/\pi)^{3/2}\exp[-\gamma_j(\mathbf{v}_i-\mathbf{v}_j)^2]\;.
\end{equation*}
On the right-hand-side of Eq.~(\ref{eq:weight_evolution}), the tensor $C_{ab}^{ik\ell}$ becomes
\begin{equation*}
  (C^{ik\ell}_{ab})_{\rm GP} =
    \int d\mathbf{v} f_i(\mathbf{v})\,C^{k\ell}_{ab}(\mathbf{v}) \; ,
    \quad (C^{ik\ell}_{ab})_{\rm CC} = C^{k\ell}_{ab}(\mathbf{v}_i) \;,
\end{equation*}
for the Galerkin and center-collocation, respectively.  Obtaining the center-collocation tensor
$(C^{ik\ell}_{ab})_{\rm CC}$ is merely a matter of evaluating the RBF tensor $C^{k\ell}_{ab}(\mathbf{v})$
at the collocation points.  Evaluation of the tensor $(C^{ik\ell}_{ab})_{\rm GP}$ for the Galerkin
projection is somewhat more complicated, although the
result may be potentially be more accurate or robust.  Nevertheless, to maintain simplicity,
we focus hereafter on the center collocation-method and omit the \textrm{CC} subscript for brevity.
In this case the RBF equations for the full nonlinear system become
\begin{equation}
\label{eq:rbf_cc_kinetic}
\sum_{j} M_{ij} \, {\cal L}_{ij} w_a^j
 = \sum_{k,\ell,b} w_{a}^{k} \, w_{b}^{\ell}\,C^{ik\ell}_{ab}  ,\quad\forall\quad i\in\,1,2,3,\dots,
\end{equation}
where the operator ${\cal L}$ 
\begin{equation*}
{\cal L}_{ij} \doteq \frac{\partial}{\partial t} + \mathbf{v}_i\cdot\nabla +
\frac{e_a}{\trbf_j^{a}} \Big[ 
(\mathbf{v}_j-\mathbf{v}_i) \cdot \mathbf{E} + (\mathbf{v}_j \times \mathbf{v}_i) \cdot \mathbf{B} \Big] 
\end{equation*}
retains the familiar appearance of the Vlasov operator even though the velocity-space has been
completely removed from the problem.  Note that ${\cal L}$ depends explicitly on species $a$ and
implicitly on $b$ via the Maxwell equations.  In ${\cal L}$ we have defined the temperature-like
parameter $\trbf_i^{a}=m_a/(2\gamma_i)$ and also dropped some additional terms that arise if the
parameters $\gamma_i$ and $\mathbf{v}_i$ depend on position.  The RBF-kinetic equation~(\ref{eq:rbf_cc_kinetic})
describes collisional fluid-like evolution of the weights in time and space.  One physically appealing
feature of the RBF expansion is that familiar expressions for macroscopic fluid moments retain
their intuitive form:
\begin{align*}
\textrm{Density}~~~n_a &~= \sum_i w_{a}^{i} \\
\textrm{Velocity}~~~n_a \mathbf{V}_a &~= \sum_i w_{a}^{i} \, \mathbf{v}_i \\
\textrm{Temperature}~~~\frac{3}{2} n_a T_a &~= \sum_i w_{a}^{i}
\left[ \frac{3}{2} \trbf_i^a + \frac{1}{2} m_a (\mathbf{v}_i-\mathbf{V}_a)^2 \right] \\
\textrm{Momentum flux tensor}~~~ \mathbf{\Pi}_a  &~= \sum_i w_{a}^{i} \, m_a 
\Big[ \trbf_i^a \, \mathbb{I} + \mathbf{v}_i \mathbf{v}_i \Big]
\\
\textrm{Energy flux}~~~\mathbf{Q}_a &~= \sum_i w_{a}^{i} \, \mathbf{v}_i
\left[ \frac{5}{2} \trbf_i^a + \frac{1}{2} m_a v_i^2 \right], 
\end{align*}
Another is that Maxwell's equations can also written compactly:
\begin{align*}
  \nabla \cdot \mathbf{E} &~= 4 \pi \sum_a e_a n_a = 4 \pi \sum_{a,i} e_a w_{a}^{i} \; , \\
  \nabla \times \mathbf{B} &~= 4 \pi \sum_a e_a n_a \mathbf{V}_a = 4 \pi \sum_{a,i} e_a  w_{a}^{i} \mathbf{v}_i \; ,
\end{align*}  
where for simplicity we have neglected the displacement current.
In contrast to the standard fluid approach, where calculation of the higher-order
velocity-space moments becomes increasingly cumbersome, the RBF-kinetic approach, however, offers a
straightforward and intuitive alternative by describing the velocity space with Gaussian
functions that naturally appear in the kinetic theory as thermodynamic equilibrium solutions.
Moreover, so long as the RBF spacing is equal to about one e-folding length, the problem
remains well-conditioned even for hundreds or thousands of basis functions. 

It should also be evident that the Gaussian RBF method is well-suited to \textit{linearized}
models.
The linearized
collision operator in this case is just the sum of $C^{0i}$ and $C^{i0}$; namely, the
first row and column of the nonlinear collision tensor.  Because the method is mesh-free,
it is also natually suited to multi-species plasmas, unlike contemporary algorithms that
are based on a velocity mesh \cite{xiong:2008}.

\begin{figure}
\begin{center}
\includegraphics[width=0.35\textwidth]{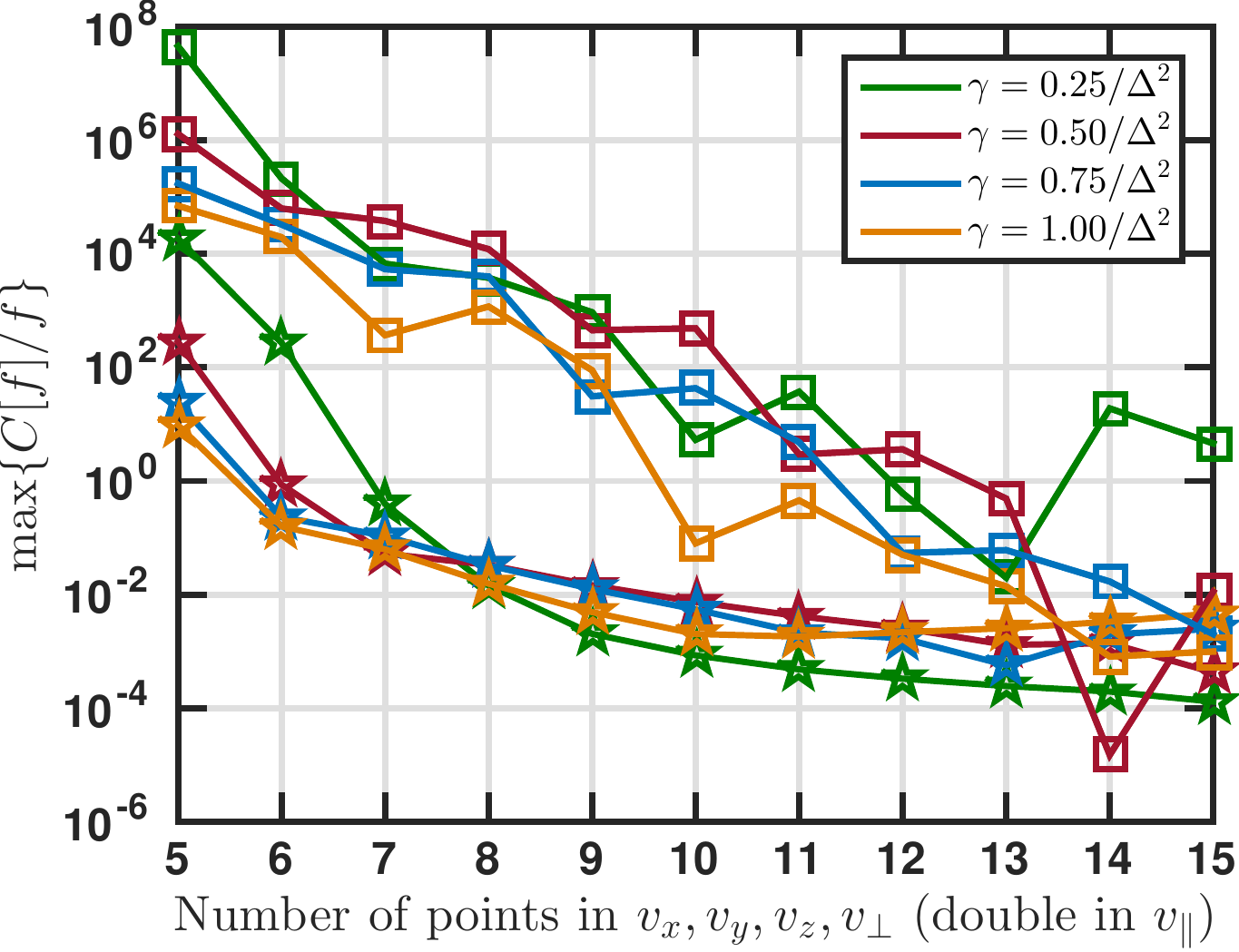}
\caption{Convergence of a steady-state collision operator. The lines with square (star) markers represent the 3D (axisymmetric 2D) solutions and $\Delta$ is the distance between adjacent collocation points.}
\label{fig:zero_test}
\end{center}
\end{figure}

{\bf Nonlinear Simulations in 2D and 3D}~~ 
To assess the viability of the Gaussian RBF approach, we study a single species plasma
using a uniform grid for the collocation points $\mathbf{v}_i$ and a fixed value
$\gamma_i = \gamma$ for all RBFs.  We also normalize time with the collision time, i.e.,
$\tau=L_{aa}t$. Our problem is thus to solve the nonlinear single species Fokker-Planck
equation
\begin{equation}
\label{eq:test_equation}
\frac{\partial f}{\partial\tau} = f^2 -
\frac{\partial^2\psi}{\partial\mathbf{v}\partial\mathbf{v}}
\hskip -1pt : \hskip -1pt
\frac{\partial^2 f}{\partial\mathbf{v}\partial\mathbf{v}}\equiv C[f]
\end{equation}
We emphasize that this is a highly nontrivial numerical problem.  First, we test 
how well the discretization preserves a (stationary) Maxwellian equilibrium state,
where the temperature of the equilibrium is fixed and much warmer than the individual
RBFs.  For the analytic equilibrium we choose
$f_M=\left(\beta/\pi\right)^{3/2}\exp\left[-\beta\mathbf{v}^2\right]$, with $\beta=1/4$
and arrange the collocation points into uniform rectilinear grids in spaces
$[v_x,v_y,v_z]\in[-6.5,6.5]\times[-6.5,6.5]\times[-6.5,6.5]$ and $[v_{\parallel},v_{\perp}]\in[-6.5,6.5]\times[0,6.5]$ for the full 3D and axisymmetric 2D methods respectively.
The equilibrium is projected onto
the RBF basis to obtain the numerical equivalent of the steady state distribution
function.  Then the collision operator in Eq.~(\ref{eq:test_equation}) is evaluated for
an increasing number of collocation points and different values of~$\gamma$.  As
an error metric, we choose $\max \{ C[f]/f \}$, presented in Fig.~\ref{fig:zero_test}.
This test is sensitive to the error both in the region interior to the RBF collocation centers,
as well as to the region beyond them.  As is evident from
Fig.~\ref{fig:zero_test}, the maximum of the global value decreases when the
number of collocation points is increased indicating that a numerical equivalent to the analytical steady-state can be reached.

\begin{figure}
\begin{center}
\includegraphics[width=0.35\textwidth]{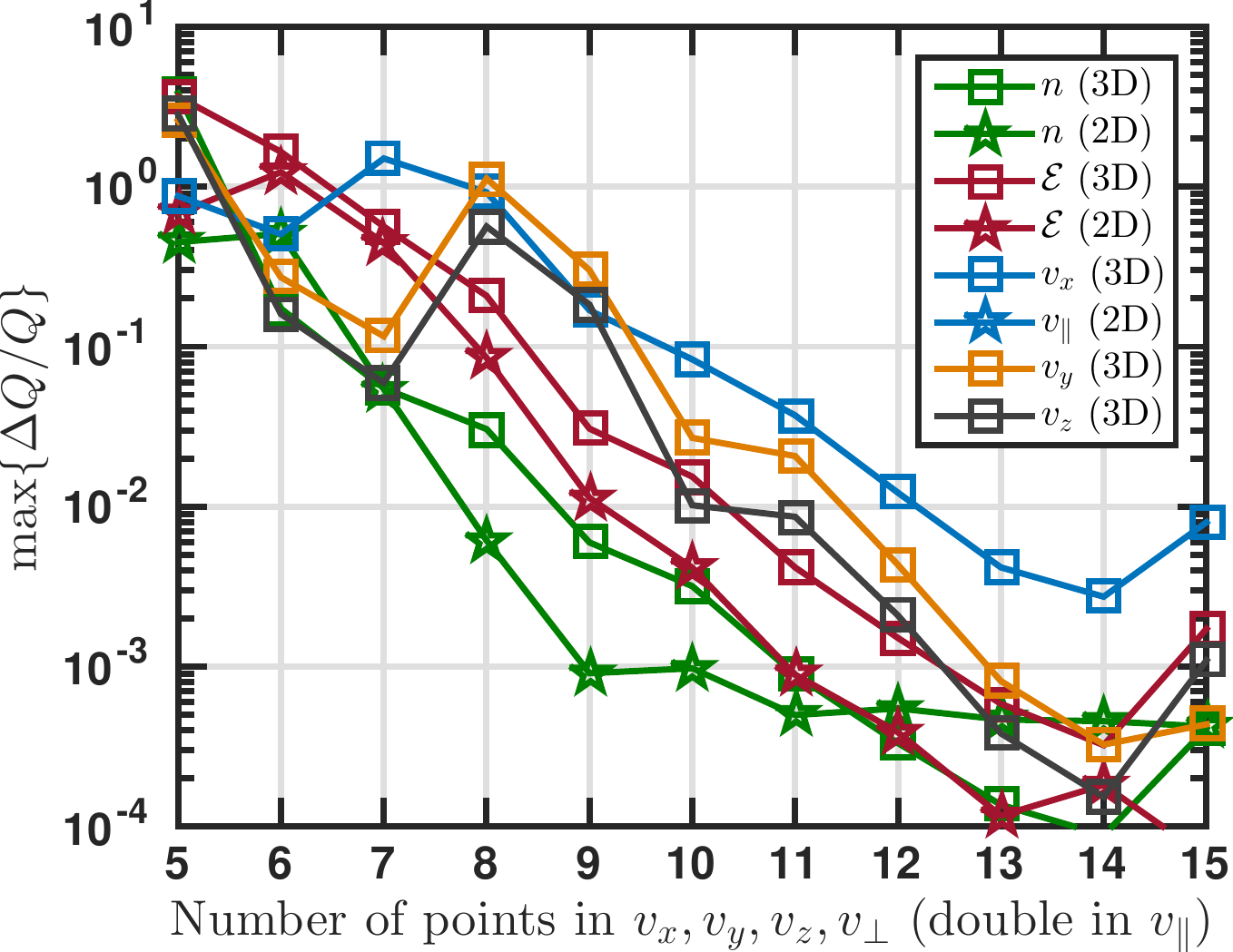}
\caption{Convergence of the velocity space moments in a relaxation study. Here $\gamma=0.5/\Delta^2$ was used and $\Delta$ is the distance between adjacent collocation points.}
\label{fig:conservation_test}
\end{center}
\end{figure}

Next, we follow the relaxation of a non-trivial distribution function to the
equilibrium state and track the conservation of number, momentum, and energy densities.
The initial state of the distribution function is set to $f(\mathbf{v},\tau=0)=\sum_{i=1}^{2}e^{-\beta_i(\mathbf{v}-\mathbf{v}_i)^2}$,
where $\beta_i=1/5$, $\mathbf{v}_1=(3,0,0)$, $\mathbf{v}_2=(-3,0,0)$ for the $(v_x,v_y,v_z)$ components and respectively. The initial state is thus axisymmetric with respect to the axis $v_x$ and the the axisymmetric setting is thus chosen to be $(v_{\parallel}=v_x,v_{\perp}^2=v_z^2+v_y^2)$. The collocation points are chosen uniformly in the regions $[v_x,v_y,v_z]\in[-8,8]\times[-8,8]\times[-8,8]$ and $[v_{\parallel},v_{\perp}]\in[-12,12]\times[0,12]$, the initial state projected to the Gaussian RBF basis to get initial values for the weights, and the weights then evolved in time according to the Eq.~(\ref{eq:weight_evolution}) using a standard fourth-order Runge-Kutta method. 
From the time-evolution of the weights, we have calculated the evolution of the velocity-space moments, and recorded their maximal deviations from the initial values during the simulation. The results are illustrated in Fig.~\ref{fig:conservation_test}, where 
one observes good conservation of number ($n$), momentum (velocity components $v_x$, $v_y$, and $v_z$), and energy ($\mathcal{E}$) densities, and convergence of their relative error towards zero when the number of RBFs is increased for both full 3D and axisymmetric 2D solvers. Even better conservation properties can be reached by embedding the conservation laws into the time-evolution of the weights. For example, replacing one row in the matrix $(M_{ij})_{\rm CC}$ with ones (unity moments of the basis functions) and setting the corresponding element in $C_{ikl,CC}$ to zero neglects information from one of the basis nodes but adds the density conservation. With this approach we observed, e.g., density conservation up to 12 digits without relevant increase in the computation time.

\newcommand\fwidth{.205\textwidth}
\begin{figure*}[!t]
  \centering
  \subfigure[\quad $\tau=0$]{
    \centering
    \includegraphics[width=\fwidth]{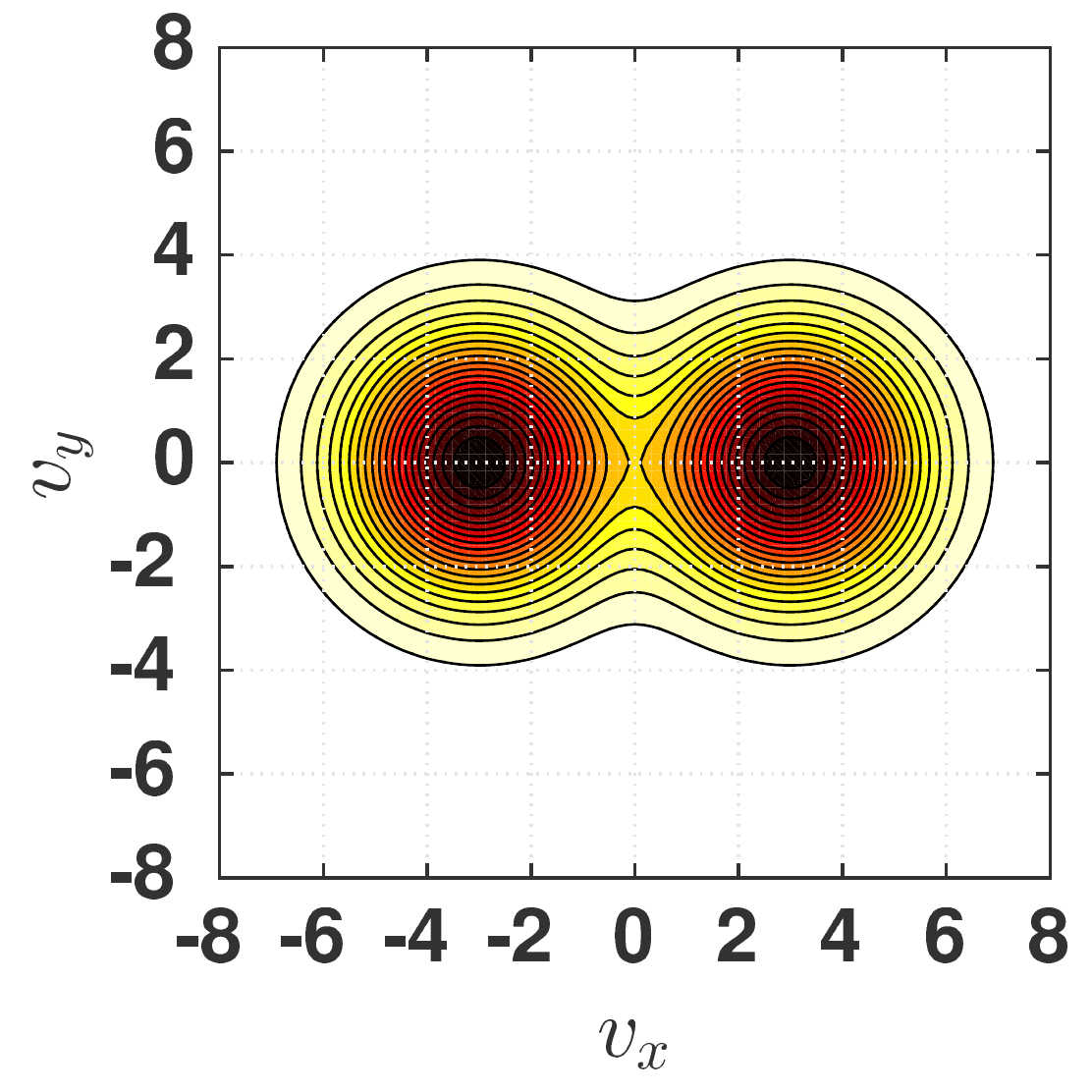}
    \label{fig:evolution_test_a}
  }
  \subfigure[\quad $\tau=3$]{
    \centering
    \includegraphics[width=\fwidth]{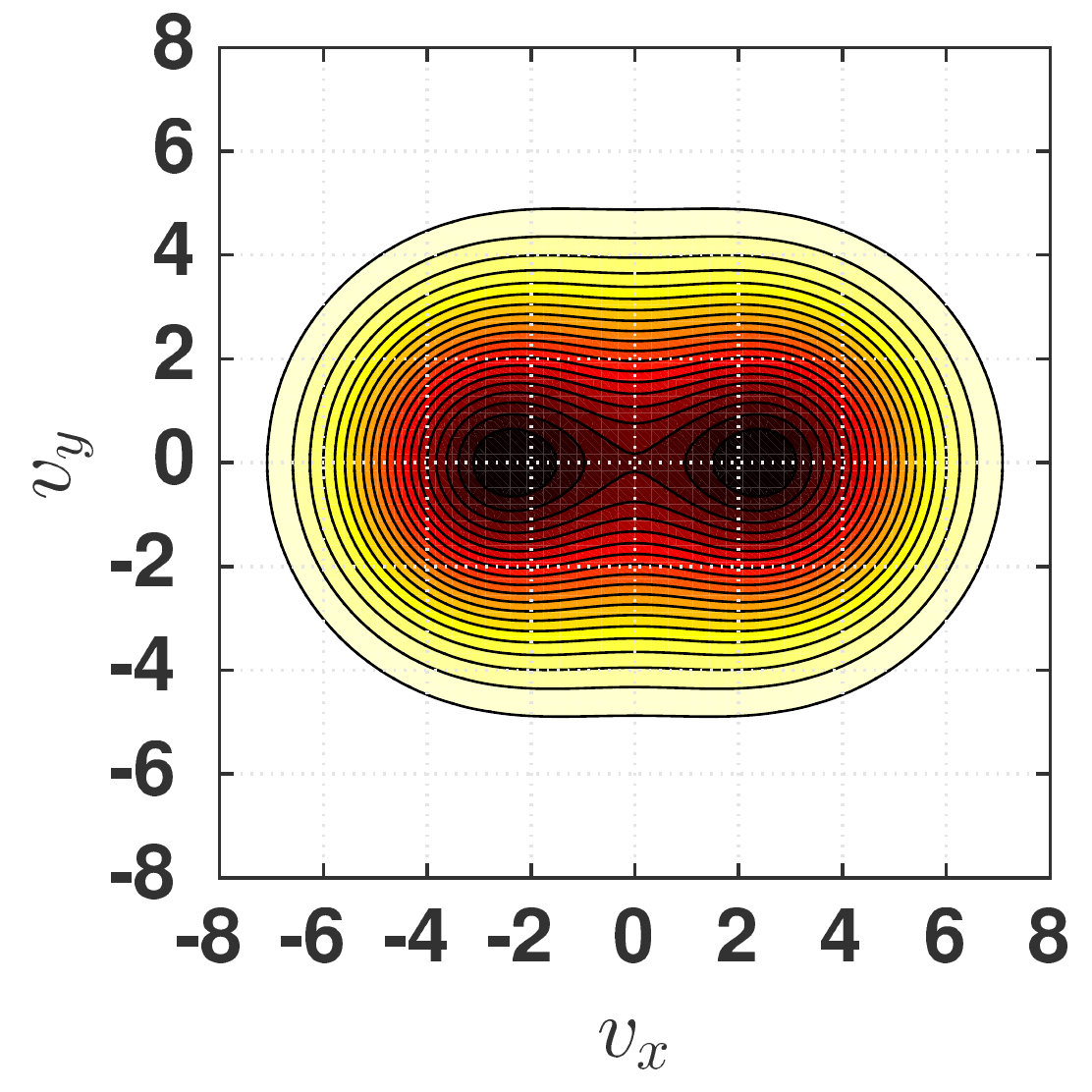}
    \label{fig:evolution_test_b}
  }
  \subfigure[\quad $\tau=6$]{
    \centering
    \includegraphics[width=\fwidth]{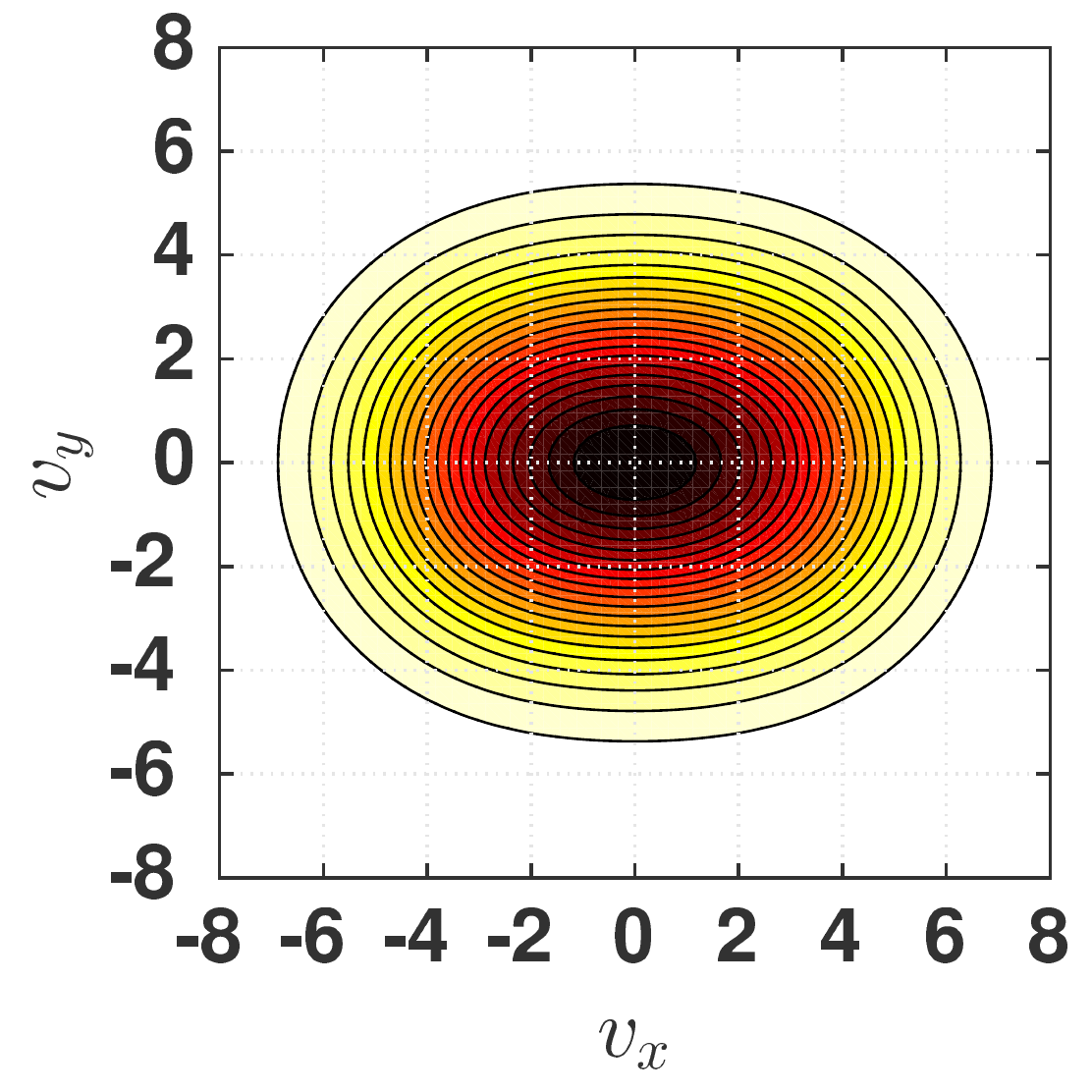}
    \label{fig:evolution_test_c}
  }
  \subfigure[\quad $\tau=40$]{
    \centering
    \includegraphics[width=\fwidth]{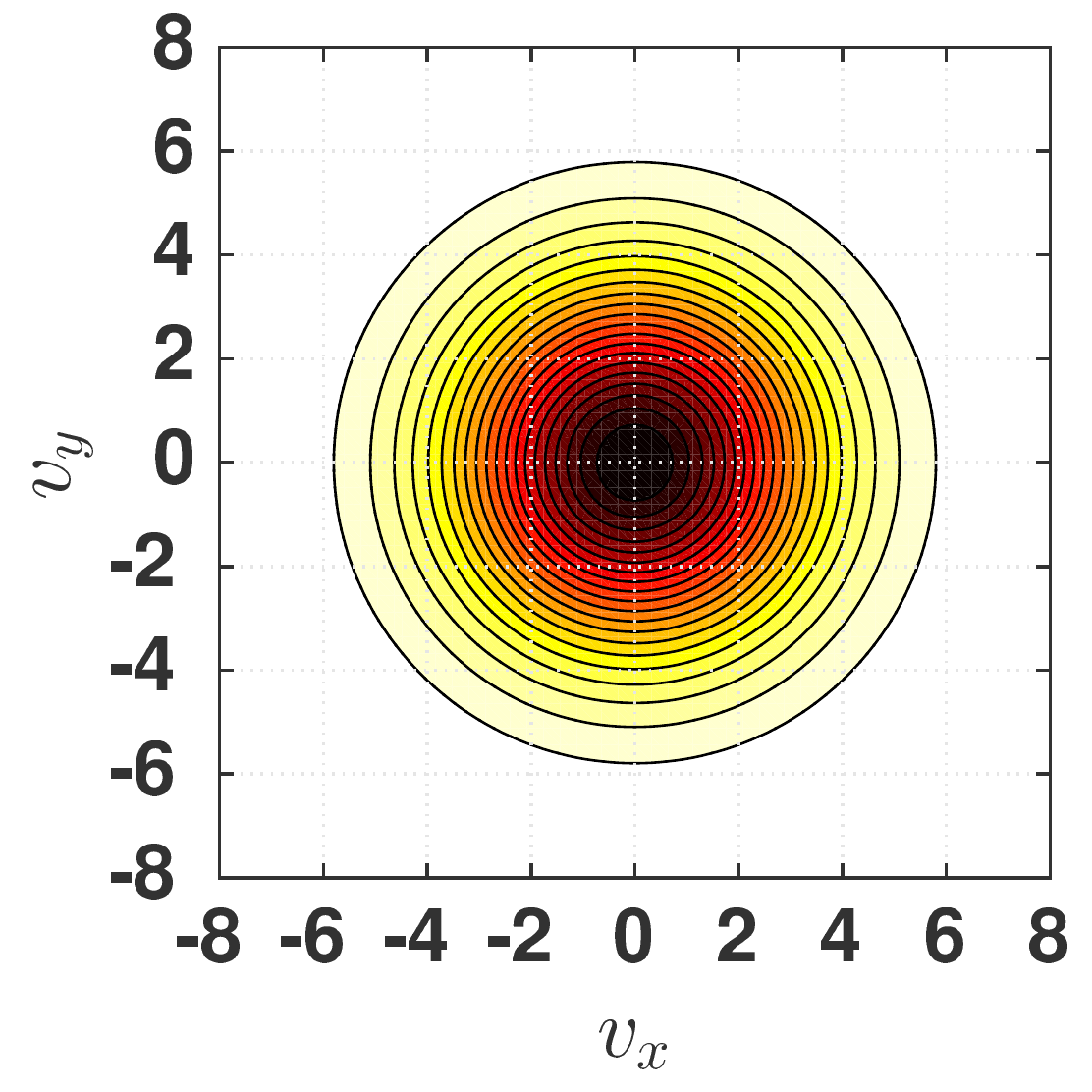}
     \label{fig:evolution_test_d}
   }
  \subfigure[\quad $\tau=0$]{
    \centering
    \includegraphics[width=\fwidth]{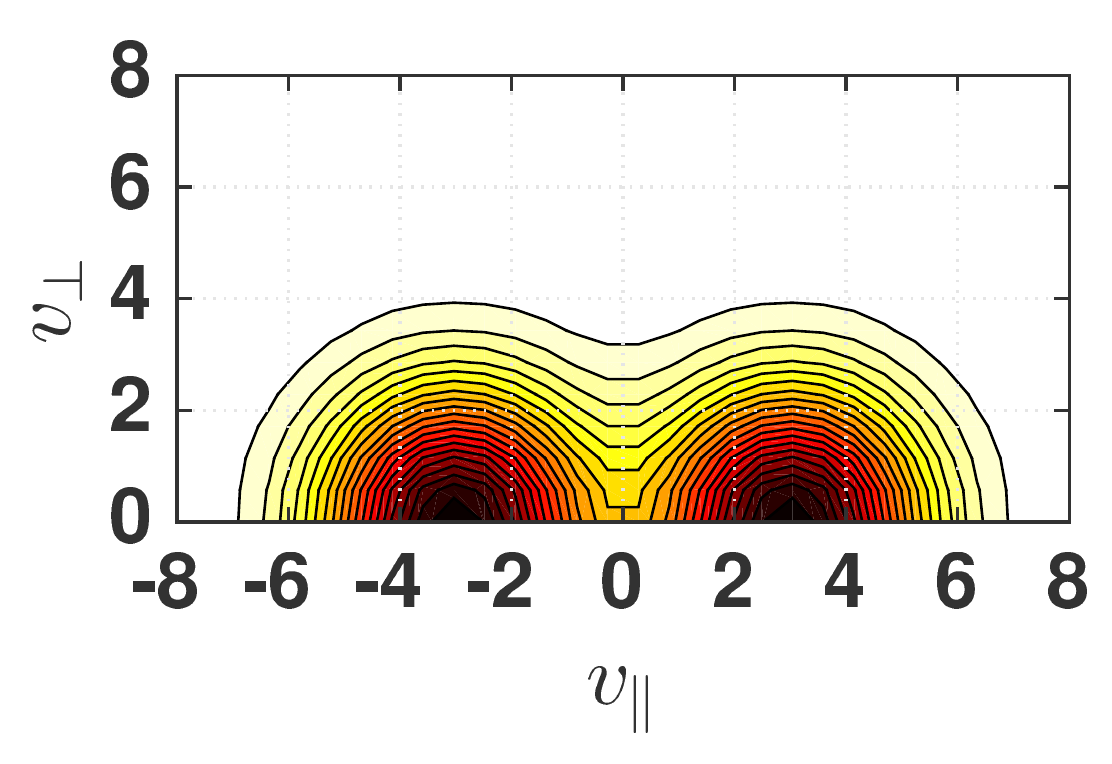}
    \label{fig:evolution_test_e}
  }
  \subfigure[\quad $\tau=3$]{
    \centering
    \includegraphics[width=\fwidth]{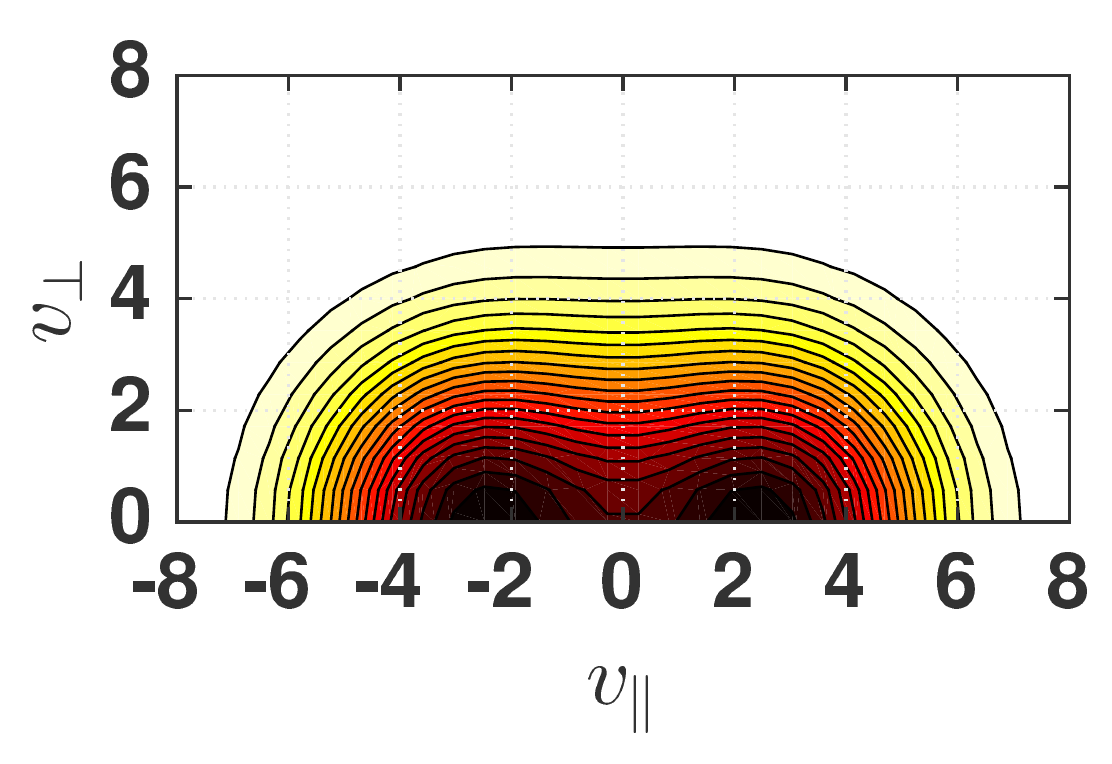}
    \label{fig:evolution_test_f}
  }
  \subfigure[\quad $\tau=6$]{
    \centering
    \includegraphics[width=\fwidth]{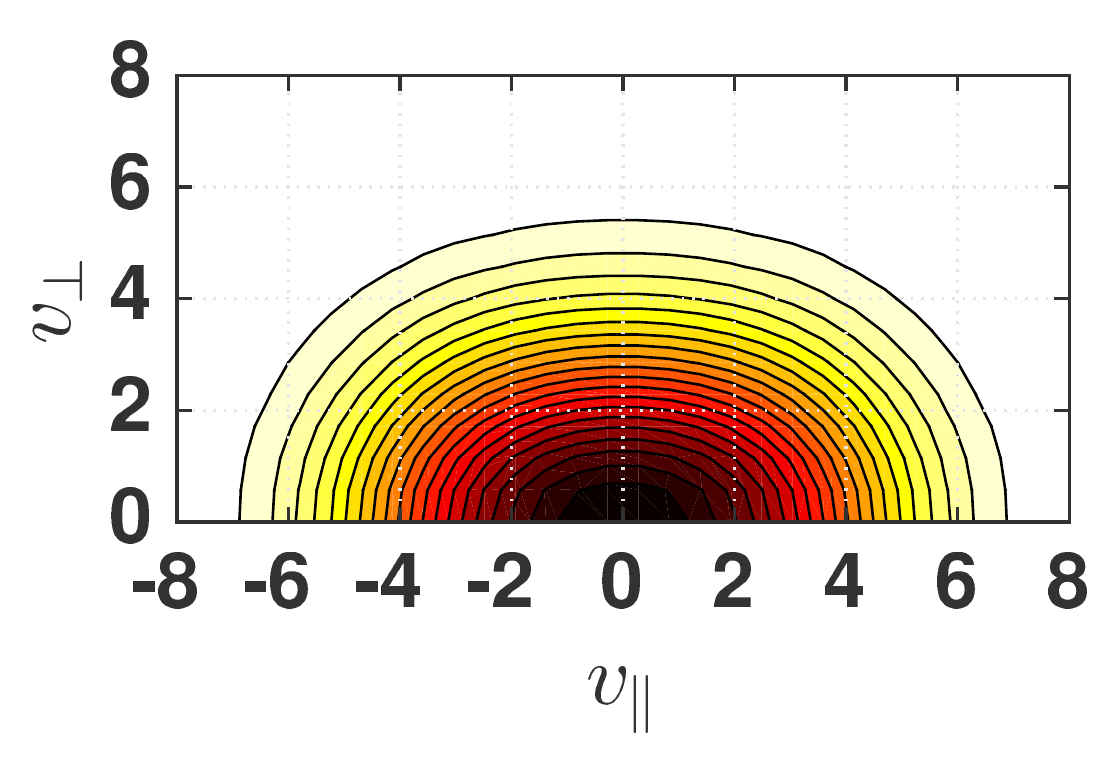}
    \label{fig:evolution_test_g}
  }
  \subfigure[\quad $\tau=40$]{
    \centering
    \includegraphics[width=\fwidth]{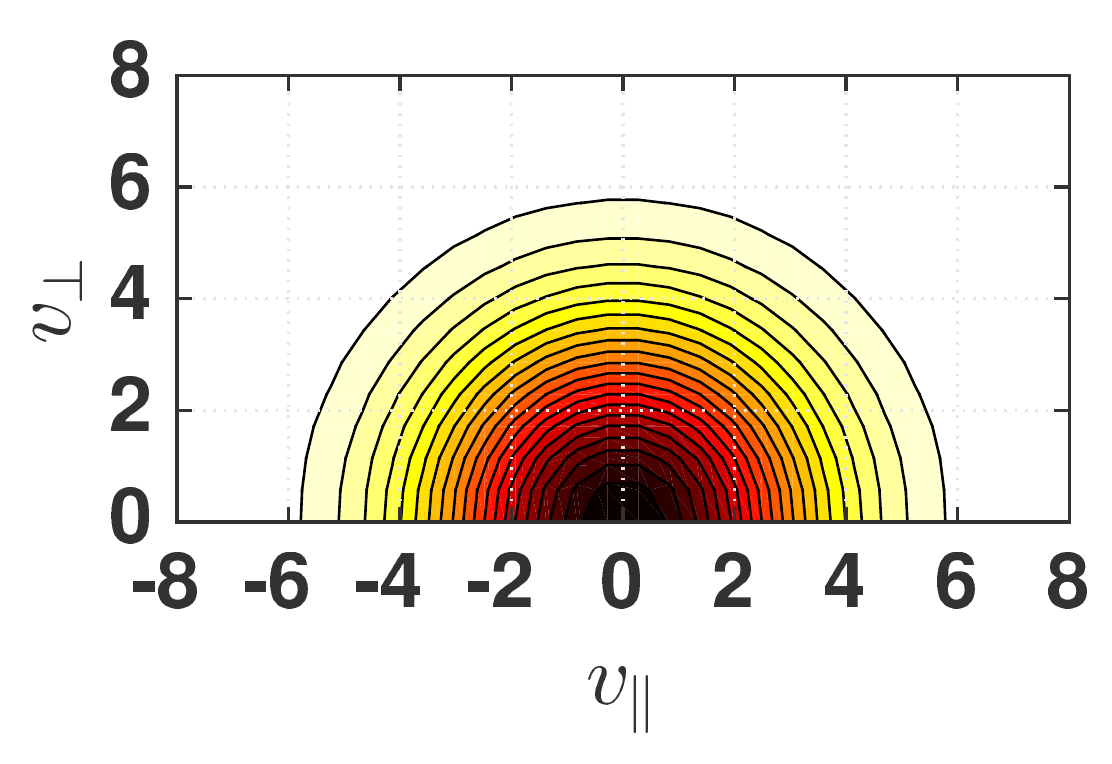}
     \label{fig:evolution_test_h}
   }
   \caption{Relaxation of an axisymmetric initial state towards spherically symmetric equilibrium state from the 3D solver (a,b,c,d) and axisymmetric 2D solver (e,f,g,h). The solution is calculated using $15\times15\times15$ 3D RBFs and $30\times15$ axisymmetric RBFs. }
   \label{fig:evolution_test}
\end{figure*}
\begin{figure}[!h]
\begin{center}
\includegraphics[width=0.35\textwidth]{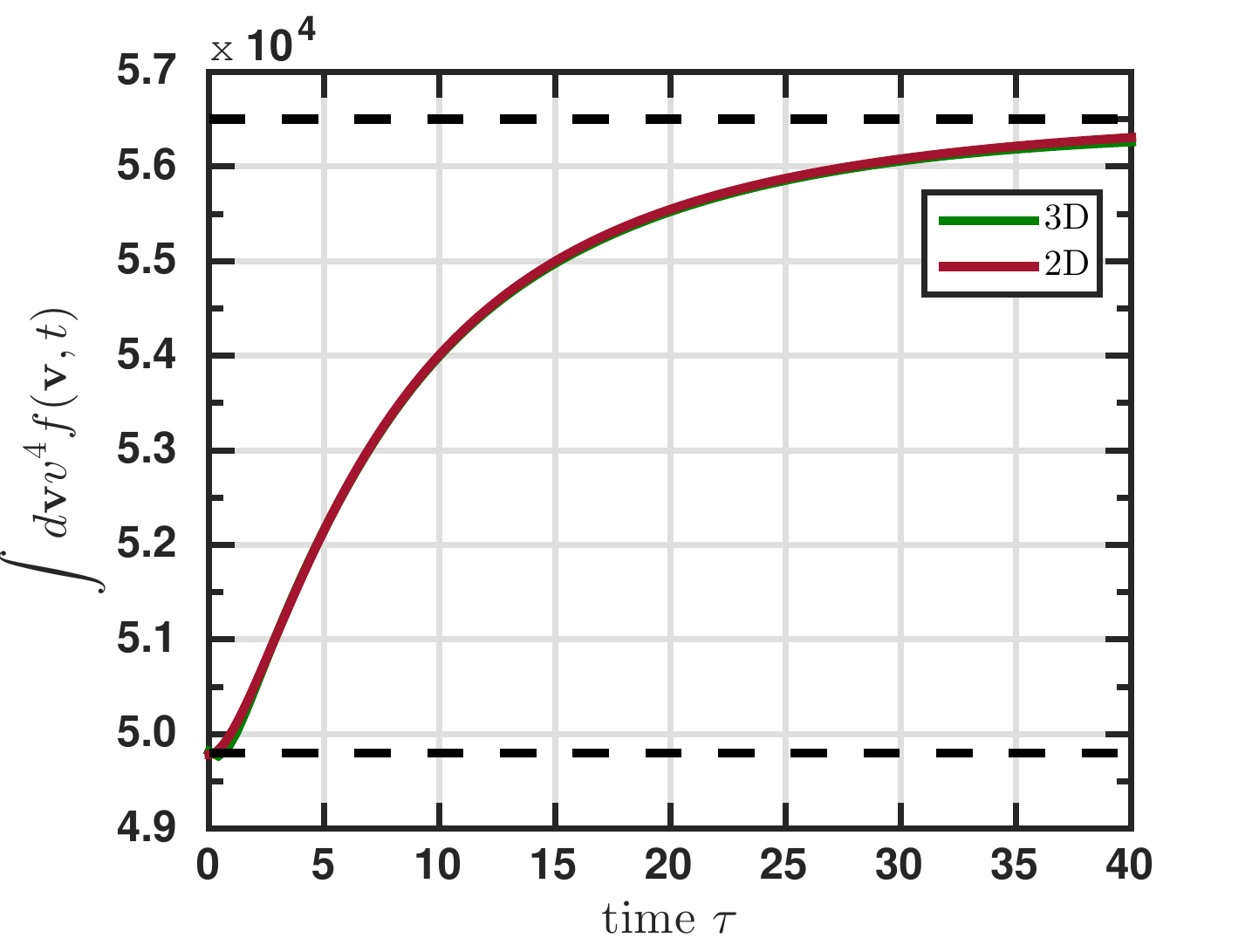}
\caption{Time evolution of the $v^4$-moment during the relaxation study (the almost overlapping red and green curves). The lower (upper) dashed line represents the analytical initial (equilibrium) values.}
\label{fig:v4}
\end{center}
\end{figure}

The time evolution of the distribution in the conservation study using $15^3$ 3D RBFs and $30\times15$ axisymmetric RBFs is illustrated in Fig.~\ref{fig:evolution_test} with slices in $(v_x,v_y)$-plane for the full 3D solver and in $(v_{\parallel},v_{\perp})$-plane for the 2D axisymmetric solver. The time slices are chosen for the initial state ($\tau=0$) in Figs.~\ref{fig:evolution_test_a} and~\ref{fig:evolution_test_e}, for the beginning of the relaxation process ($\tau=3$) in Figs.~\ref{fig:evolution_test_b} and~\ref{fig:evolution_test_f}, for the merging phase towards the equilibrium ($\tau=6$) in Figs.~\ref{fig:evolution_test_c} and~\ref{fig:evolution_test_g}, and for the final state ($\tau=40$) where the distribution function is close to spherical symmetric and equilibrium in Figs.~\ref{fig:evolution_test_d} and~\ref{fig:evolution_test_h}. Both the 2D and 3D solvers qualitatively describe the same solution to the initially axisymmetric problem and preserve the initial symmetry.
As a more quantitative measure, we also follow the time evolution of the $v^4$-moment of the distribution function which is not a conserved quantity. From the analytical initial state one can calculate the value to be $4.98\times10^4$ and for the corresponding analytical equilibrium state the value is $5.65\times10^4$. The time evolution of $v^4$-moment from the numerical calculation is shown in Fig.~\ref{fig:v4}. As is clear, the 2D and 3D solutions agree with each other and also confirm our claim that the solution in Figs.~\ref{fig:evolution_test_d} and~\ref{fig:evolution_test_h} is close to the analytical equilibrium state. In other words, we have solved the relaxation problem accurately in both 2D and 3D and demonstrated the capabilities of the RBFs for discretizing the nonlinear collision operator.

{\bf Summary}~~
In this letter, we have derived a completely new approach to collisional dynamics in plasmas. We have shown how to implement a Gaussian radial-basis-function ansatz to discretize the velocity space, and demonstrated its capabilities in non-linear Fokker-Planck calculations in both full 3D and axisymmetric 2D velocity-space. In a wider connection to kinetic theory we have provided also the discretization of the advection terms and given analytical ansatz expressions for all relevant fluid-moments. In future, our method could be used to solve various problems where non-equilibrium kinetic effects are important, as long as the Landau-approximation for the collision integral is valid.


\providecommand{\noopsort}[1]{#1}

\end{document}